\documentclass[epj]{svjour}
\def\be{\begin{equation}}
\def\ee{\end{equation}}
\def\bea{\begin{eqnarray}}
\def\eea{\end{eqnarray}}
\def\ba{\begin{array}}
\def\ea{\end{array}}

\usepackage{graphics}
\begin{document}
\title{Revisit emission spectrum and entropy quantum of the Reissner-Nordstr\"{o}m black hole}
\author{Qing-Quan Jiang\inst{1} }
\offprints{}          
\institute{College of Physics and Electronic Information, China
 West Normal University, Nanchong, \\ Sichuan 637002, People's Republic of China}
\date{Received: date / Revised version: date}
%
\abstract{Banerjee and Majhi's recent work shows that black hole's emission spectrum could be fully reproduced in the tunneling picture, where, as an intriguing technique, the Kruskal extension was introduced to connect the left and right modes inside and outside the horizon. Some attempt, as an extension, was focused on producing the Hawking emission spectrum of the (charged) Reissner-Nordstr\"{o}m black hole in the Banerjee-Majhi's treatment. \emph{Unfortunately}, the Kruskal extension in their observation was so badly defined that the ingoing mode was \emph{classically} forbidden traveling towards the center of black hole, but could \emph{quantum} tunnel across the horizon with the probability $\Gamma=e^{-\pi \omega_0/\kappa_+}$. This tunneling picture is \emph{unphysical}. With this point as a central motivation, in this paper we first introduce such a suitable Kruskal extension for the (charged) Reissner-Nordstr\"{o}m black hole that a perfect tunneling picture can be provided during the charged particle's emission. Then, under the new Kruskal extension, we revisit the Hawking emission spectrum and entropy spectroscopy as tunneling from the charged black hole. The result shows that the tunneling method is so universally robust that the Hawking blackbody emission spectrum from a charged black hole can be well reproduced in the tunneling mechanism, and its induced entropy quantum is a much better approximation for the forthcoming quantum gravity theory.
} 
\maketitle
\section{Introduction}
The study of black hole physics has been a long-standing and hot topic in theoretical physics since the birth of Einstein's theory of gravitation. In particular,
exploring quantum properties (Hawking radiation, entropy quantum, etc) of black holes has important physical significance, which may provide a window to find an effective way to quantize gravitational field. However, a self-consistent theory of quantum gravitation was lacking so far. Hence, it may be an appropriate juncture to ``take a step back" and re-enforce our understanding of these issues at a semiclassical level.

With this in mind, a semiclassical treatment that implemented the Hawking radiation as a tunneling process was initiated by Kraus and Wilczek \cite{r3} and then developed by Parikh and Wilczek \cite{MF} (this framework shall be referred to as the Kraus-Parikh-Wilczek's analysis for briefness, see also Ref. \cite{tunneling1} for a different methodology that the tunneling picture has been
applied.). Later, the tunneling mechanism as the most fashionable model to explore the quantum properties of black holes, has received so popular attention that it can be applicable in the context of a variety of black holes \cite{t2,t3,t4}, fermion emission \cite{t5}, as well as the presence of the higher order quantum corrections while going beyond the semiclassical approximation \cite{t6} etc.

In these observations \cite{r3,MF,tunneling1,t2,t3,t4,t5,t6}, the semiclassical Hawking temperature can be easily obtained by exploiting the form of the semiclassical tunneling rate, but there was a glaring omission since all of them failed to yield directly the blackbody spectrum. To rectify this shortcomings, Banerjee and Majhi, with the aid of density matrix techniques, reformulated the tunneling mechanism to directly find the Hawking emission spectrum as tunneling from the black hole horizon \cite{c1}. Then, based on the modified version of the tunneling mechanism, the entropy spectroscopy from the black hole horizon was also described in the tunneling framework \cite{RBE}. Later, as an extension, the Banerjee-Majhi's treatment was developed in \cite{c2,BRM} to the case of black hole in nontrivial gravity, then further improved by us to include the effect of back reaction \cite{JHC} and higher order corrections \cite{JHC1}. Upon achieving great success along this line, however, it is not clear to explore the quantum properties (Hawking radiation, entropy quantum, etc) of a \emph{charged} black hole in the Banerjee-Majhi's treatment. In the work \cite{zhaor}, one attempt was focused on this issue. \emph{Unfortunately}, their treatment violated the fundamental description of black hole physics since the Kruskal extension as an illuminating technique in the Banerjee-Majhi's treatment was so badly defined in \cite{zhaor} that the ingoing mode was classically forbidden traveling towards the center of black hole, but could quantum tunnel across the horizon with the probability $\Gamma=e^{-\pi \omega_0/\kappa_+}$. Motivated by this fact, in this paper, we will revisit the Banerjee-Majhi's treatment in the context of the (charged) Reissner-Nordstr\"{o}m black hole. In doing so, it is necessary to introduce such a suitable Kruskal extension for the charged black hole that a perfect tunneling picture can be provided during the charged particle's emission. Then, under the new Kruskal extension, we will revisit the Hawking emission spectrum and entropy spectroscopy as tunneling from the charged black hole.

The organization of the paper goes as follows. In Sec. \ref{sec1}, we first review the errors of the Kruskal extension, introduced by Zhao etc., for the charged black hole, and analyze its induced unphysical phenomenon. Then, to develop the Banerjee-Majhi's treatment applicable for the charged black hole, we introduce such a suitable Kruskal extension that a perfect tunneling picture can be existed during charged particle's emission. Under the new Kruskal extension introduced by us, Sec. \ref{sec2} and \ref{sec3} are, respectively, devoted to revisit the Hawking emission spectrum and entropy spectroscopy as tunneling from the charged black hole. Sec. \ref{sec4} ends up with conclusion and discussion.

\section{The Kruskal extension for the Reissner-Nordstr\"{o}m black hole} \label{sec1}

In the Banerjee-Majhi's treatment, the Kruskal extension as an intriguing technique was introduced to connect the left and right modes inside and outside the horizon. In \cite{zhaor}, much effort was focus on developing the semiclassical treatment to the case of the (charged) Reissner-Nordstr\"{o}m black hole. However, the Kruskal extension in their observation was so badly defined for the charged black hole that the fundamental description of black hole physics was broken here. In this section, we aim to find a suitable Kruskal extension for the (charged) Reissner-Nordstr\"{o}m black hole, under which a perfect tunneling phenomenon can be provided during the charged particle's emission. Before that, Let's first review the Kruskal extension, introduced by Zhao etc., for the charged black hole, and get at the root of errors.

\subsection{The Kruskal extension by Zhao etc.}

In Ref. \cite{zhaor}, Zhao etc. tried to develop the Banerjee-Majhi's treatment to the case of the (charged) Reissner-Nordstr\"{o}m black hole. To provide a perfect tunneling picture during the charged particle's emission, they introduced a Kruskal extension to eliminate the coordinate singularity at the horizon, which was given by
\begin{eqnarray}
&&T_{\textrm{in}} = e^{\kappa_+(r_\ast)_{\textrm{in}}}\cosh(\kappa_+ t_{\textrm{in}}); \nonumber\\
&&X_{\textrm{in}} = e^{\kappa_+(r_\ast)_{\textrm{in}}}\sinh(\kappa_+ t_{\textrm{in}});~~~~~~~~(r< r_+)\nonumber\\
&&T_{\textrm{out}}=e^{\kappa_+(r_\ast)_{\textrm{out}}}\sinh(\kappa_+ t_{\textrm{out}});\nonumber\\
&&X_{\textrm{out}}=e^{\kappa_+(r_\ast)_{\textrm{out}}}\cosh(\kappa_+ t_{\textrm{out}});~~~~(r> r_+) \label{eq1}
\end{eqnarray}
where $\kappa_+$ is the surface gravity of the Reissner-Nordstr\"{o}m black hole, and $r_\ast$ is defined as $r_\ast=\int \frac{dr}{f(r)}$. Under the Kruskal extension as above, together with the mapping $T_{\textrm{in}}\rightarrow T_{\textrm{out}}$ and $X_{\textrm{in}}\rightarrow X_{\textrm{out}}$, one metric is commonly valid just inside and outside the horizon.
With this in mind, one have
\bea
&&t_{\textrm{in}}\rightarrow t_{\textrm{out}}-i\frac{\pi}{2\kappa_+};~~(r_\ast)_{\textrm{in}}\rightarrow (r_\ast)_{\textrm{out}}+i\frac{\pi}{2\kappa_+}. \label{eq2}
\eea
From this mapping, one can connect the left and right modes defined inside and outside the horizon. In \cite{zhaor}, the left and right modes inside and outside the horizon of the Reissner-Nordstr\"{o}m black hole are defined by
\begin{eqnarray}
&&\Phi_{\textrm{in}}^{R}=e^{-\frac{i}{\hbar}\omega v_{\textrm{in}}}; ~~~~~~~ \Phi_{\textrm{out}}^{R}=e^{-\frac{i}{\hbar}\omega v_{\textrm{out}}};\nonumber\\
&&\Phi_{\textrm{in}}^{L}=e^{-\frac{i}{\hbar}\omega u_{\textrm{in}}};~~~~~~~\Phi_{\textrm{out}}^{L}=e^{-\frac{i}{\hbar}\omega u_{\textrm{out}}}, \label{eq3}
\end{eqnarray}
with the null coordinates
\begin{equation}
u=t-\hat{r},~~~v=t+\hat{r}, \label{eq4}
\end{equation}
where $\hat{r}=\frac{\omega-\omega_0}{\omega}r_\ast$, and $\omega_0=e\frac{Q}{r_+}$ is electrostatic potential on
surface $r_+$. Hence, following the definition (\ref{eq2}) and (\ref{eq4}) one obtain the relations connecting the
null coordinates defined inside and outside the horizon,
\begin{eqnarray}
&&u_{\textrm{in}}=t_{\textrm{in}}-\hat{r}_{\textrm{in}}\rightarrow u_{\textrm{out}}-i\frac{2\omega-\omega_0}{\omega}\frac{\pi}{2\kappa_+},\nonumber\\
&&v_{\textrm{in}}=t_{\textrm{in}}+\hat{r}_{\textrm{in}}\rightarrow v_{\textrm{out}}-i\frac{\omega_0}{\omega}\frac{\pi}{2\kappa_+}. \label{eq5}
\end{eqnarray}
Now, it is easy to see that under the relations (\ref{eq5}), the inside and outside modes are connected by
\begin{eqnarray}
&&\Phi_{\textrm{in}}^{R}\rightarrow e^{-{\pi(2\omega-\omega_0)}/{(2\kappa_+)}}\Phi_{\textrm{out}}^{R}, \nonumber\\
&&\Phi_{\textrm{in}}^{L}\rightarrow e^{-{\pi \omega_0}/{(2\kappa_+)}}\Phi_{\textrm{out}}^{L}.\label{eq6}
\end{eqnarray}
This connection is crucial for the Banerjee-Majhi's treatment, by which the Hawking emission spectrum and entropy spectroscopy as tunneling from the horizon could be reproduced, as that in \cite{c1,RBE}. In view of this fact, it is necessary to check whether this connection is self-consistent with the physical requirement. Next, we will review this issue. For the left moving mode, its tunneling behavior is towards the center of the black hole, so its probability to go inside is
\begin{equation}
\Gamma^{L}=|{\Phi_{\textrm{in}}^{L}}|^2\rightarrow |e^{-{\pi \omega_0}/{(2\kappa_+)}}{\Phi_{\textrm{out}}^{L}}|^2=e^{-\pi \omega_0/\kappa_+},\label{eq7}
\end{equation}
where we have used the connection (\ref{eq6}) to recast $\Phi_{\textrm{in}}^{L}$ in terms of $\Phi_{\textrm{out}}^{L}$ since measurements are done by an outside observer. On the other hand, since the right moving mode tunnels outwards the horizon, its tunneling rate is
\begin{equation}
\Gamma^{R}=|\Phi_{\textrm{in}}^{R}|^2\rightarrow |e^{-{\pi(2\omega-\omega_0)}/{(2\kappa_+)}}\Phi_{\textrm{out}}^{R}|^2=e^{-{\pi(2\omega-\omega_0)}/{\kappa_+}}.\label{eq8}
\end{equation}
After analyzing the induced result (\ref{eq7}) and (\ref{eq8}) as a result of the Kruskal extension (\ref{eq1}), one easily find the ingoing mode is \emph{classically} forbidden traveling towards the center of black hole, but can \emph{quantum} tunnel across the horizon with the probability $\Gamma=e^{-\pi \omega_0/\kappa_+}$. This realization obviously violates the fact that the left moving (ingoing) mode is expected to trap inside the black hole, and its probability to go inside, as measured by an external observer, is to be unity. This tunneling picture is \emph{unphysical}. Hence, in \cite{zhaor}, the Kruskal extension introduced by Zhao etc. failed to develop the Banerjee-Majhi's treatment to the case of the (charged) Reissner-Nordstr\"{o}m black hole black hole. In doing so, it is necessary to introduce a suitable Kruskal extension to present a perfect tunneling phenomenon during charged particle's emission.
In the next subsection, a suitable Kruskal extension for the charged black hole will be introduced by us.

\subsection{The Kruskal extension by us}

In this subsection, we focus on the central motivation of this paper, aiming to find a suitable Kruskal extension for the (charged) Reissner-Nordstr\"{o}m black hole to connect the left and right moving modes inside and outside the horizon. Before that, it is necessary to define the left and right moving modes. In the background of the Reissner-Nordstr\"{o}m black hole, by solving the usual semiclassical Hamilton-Jacobi equation with the WKB ansatz, we have
\begin{equation}
\Phi=e^{-i\omega t\pm i(\omega-\omega_0)r_\ast},\label{eq9}
\end{equation}
where the +(-) solutions denote the right(left) movers. For further discussions, it is convenient to introduce the sets of null tortoise coordinates. In \cite{zhaor}, the null coordinates were given by (\ref{eq4}). Here, we redefine them as
\begin{equation}
u=\hat{t}-r_\ast,~~~~~~ v=\hat{t}+r_\ast, \label{eq10}
\end{equation}
where $\hat{t}=\frac{\omega} {\omega-\omega_0}t$. Noted that both the null coordinates (\ref{eq4}) and (\ref{eq10}) are well defined for the Reissner-Nordstr\"{o}m black hole. In \cite{zhaor}, an undesired result comes from the incorrect choice of the Kruskal extension, rather than that of the null coordinates.
Under the null coordinates (\ref{eq10}), the left and right modes defined inside and outside the horizon can be rewritten as
\begin{eqnarray}
&&\Phi_{\textrm{in}}^{R}=e^{-\frac{i}{\hbar}(\omega-\omega_0) v_{\textrm{in}}}; ~~~~\Phi_{\textrm{out}}^{R}=e^{-\frac{i}{\hbar}(\omega-\omega_0) v_{\textrm{out}}};\nonumber\\
&&\Phi_{\textrm{in}}^{L}=e^{-\frac{i}{\hbar}(\omega-\omega_0) u_{\textrm{in}}};~~~~\Phi_{\textrm{out}}^{L}=e^{-\frac{i}{\hbar}(\omega-\omega_0) u_{\textrm{out}}}. \label{eq11}
\end{eqnarray}
Here, the left and right moving modes are defined as: if the eigenvalue of the radial momentum operator $\hat{p}(r)$, while acting on the specific solution (\ref{eq9}), is positive, the mode is right moving. Similarly, the left moving mode corresponds to a negative eigenvalue.

Next, we will introduce a suitable Kruskal extension for the charged black hole to connect the left and right mode inside and outside the horizon.
In the tunneling picture, a pair of virtual particles is created near the horizon as a result of quantum vacuum fluctuation. According to
this scenario, a pair of particles is spontaneously created just inside the horizon, the positive energy particle then tunnels out to
the infinity, and the negative energy ``partner'' remains behind and effectively lowers the mass of the black hole. This tunneling picture
can be depicted in another manner, that is, a particle/antiparticle pair is created just outside the horizon, the negative energy particle tunnels into the horizon because the negative energy orbit exists only inside the horizon, the positive energy partner is left outside and emerges at infinity. In this paper, we adopt the first depiction without loss of universality. According to this scenario, particles must tunnel across the horizon, which yields the change of the coordinate nature. It is plausible to introduce the coordinates which are viable on both sides of the horizon. Fortunately, the Kruskal coordinates are capable of acting as this actor. Next, we aim to find this Kruskal extension for the Reissner-Nordstr\"{o}m black hole, under which the coordinates are viable on both sides of the horizon.

\textbf{Case I:} Under the null coordinates (\ref{eq10}), the $(r-t)$ sector of the Reissner-Nordstr\"{o}m spacetime is transformed  as
\be
ds^2=\frac{f(r)}{4\overline{\omega}^2}\big[(1-\overline{\omega}^2)(dv+du)^2+4\overline{\omega}^2dvdu\big], \label{eq12}
\ee
where $\overline{\omega}\equiv \frac{\omega}{\omega-\omega_0}$ and $f(r)=1-\frac{2M}{r}+\frac{Q^2}{r^2}$. First, we consider the line element which is viable outside the horizon. Outside the horizon ($r> r_+$), where the observer is present, introducing the Kruskal coordinates
\begin{equation}
V_{\textrm{out}}=e^{\kappa_+ v_{\textrm{out}}}, ~~~~~~U_{\textrm{out}}=-e^{-\kappa_+ u_{\textrm{out}}},\label{eq13}
\end{equation}
and
\begin{equation}
T_{\textrm{out}}=\frac{V_{\textrm{out}}+U_{\textrm{out}}}{2}, ~~~X_{\textrm{out}}=\frac{V_{\textrm{out}}-U_{\textrm{out}}}{2},\label{eq14}
\end{equation}
we find the metric (\ref{eq12}) in the Kruskal coordinates is rewritten as
\bea
ds^2&=&\frac{f(r)}{\kappa_+^2\overline{\omega}^2}\big[(1-\overline{\omega}^2)e^{-4\kappa_+(r_\ast)_{\textrm{out}}} \nonumber\\
&\times& (X_{\textrm{out}}dT_{\textrm{out}}-T_{\textrm{out}}dX_{\textrm{out}})^2 \nonumber\\
&+&\overline{\omega}^2 e^{-2\kappa_+(r_\ast)_{\textrm{out}}}(dT_{\textrm{out}}^2-dX_{\textrm{out}}^2)\big]. \label{eq15}
\eea
In our case, to connect the right and left moving modes inside and outside the horizon, it is enough to consider the behavior of the spacetime in a very narrow region just inside and outside the horizon. Thus, near the horizon, $f(r)$ can be expanded as
\be
f(r)=f'(r_+)(r-r_+)+\frac{f''(r_+)}{2}(r-r_+)^2+\mathcal{O}(r-r_+)^3.\label{eq16}
\ee
Also, $(r_\ast)_{\textrm{out}}$ can be integrated just outside the horizon to yield
\be
(r_\ast)_{\textrm{out}}=\int \frac{dr}{f(r)}=\frac{1}{f'(r_+)}\ln \Big[\frac{(r-r_+)}{f'(r_+)+(r-r_+)f''(r_+)/2}\Big]. \label{eq17}
\ee
Here, we neglect the contributions from the higher-order terms of $\mathcal{O}(r-r_+)^3$. So, just outside the horizon, the metric (\ref{eq15}) in the Kruskal extension is given by
\bea
ds^2&=&\frac{[f'(r_+)+(r-r_+)f''(r_+)/2]^2}{\kappa_+^2\overline{\omega}^2}\big[\overline{\omega}^2 (dT_{\textrm{out}}^2-dX_{\textrm{out}}^2)\nonumber\\
&+&(1-\overline{\omega}^2)e^{-2\kappa_+(r_\ast)_{\textrm{out}}}
(X_{\textrm{out}}dT_{\textrm{out}}-T_{\textrm{out}}dX_{\textrm{out}})^2
\big] \label{eq18}
\eea
Obviously, the new metric in the Kruskal coordinate system has no any coordinate singularity at the horizon irrespective of choosing any particular $f(r)$. Also, according to Eqs.(\ref{eq13}) and (\ref{eq14}), the Kruskal coordinates ($T_{\textrm{out}}, X_{\textrm{out}} $) which are valid outside the horizon, are given by
\bea
&&T_{\textrm{out}} = e^{\kappa_+(r_\ast)_{\textrm{out}}}\sinh(\kappa_+ \hat{t}_{\textrm{out}}), \nonumber\\
&&X_{\textrm{out}} = e^{\kappa_+(r_\ast)_{\textrm{out}}}\cosh(\kappa_+ \hat{t}_{\textrm{out}}).\label{eq19}
\eea

\textbf{Case II:}
In a similar way, inside the horizon $r< r_+$, we introduce the Kruskal coordinates as
\begin{equation}
V_{\textrm{in}}=e^{\kappa_+ v_{\textrm{in}}}, ~~~~~~U_{\textrm{in}}=e^{-\kappa_+ u_{\textrm{in}}},\label{eq20}
\end{equation}
and
\begin{equation}
T_{\textrm{in}}=\frac{V_{\textrm{in}}+U_{\textrm{in}}}{2}, ~~~X_{\textrm{in}}=\frac{V_{\textrm{in}}-U_{\textrm{in}}}{2}.\label{eq21}
\end{equation}
Here, a relative sign difference appears between the functional choice of $U_{\textrm{in}}$ and $U_{\textrm{out}}$, which ensures that the inner portion of the extended spacetime metric remains timelike. If the definition of $U_{\textrm{in}}$ follows that of $U_{\textrm{out}}$, a spacelike metric interval will be provided, which we  want to avoid since there is no coordinate singularity at the horizon. Under the Kruskal coordinates ($T_{\textrm{in}}, X_{\textrm{in}}$), the metric (\ref{eq12}) can then be rewritten as
\bea
ds^2&=&\frac{f(r)}{\kappa_+^2\overline{\omega}^2}\big[(1-\overline{\omega}^2)e^{-4\kappa_+(r_\ast)_{\textrm{in}}}(X_{\textrm{in}}dT_{\textrm{in}}-T_{\textrm{in}}dX_{\textrm{in}})^2 \nonumber\\
&-&\overline{\omega}^2 e^{-2\kappa_+(r_\ast)_{\textrm{in}}}(dT_{\textrm{in}}^2-dX_{\textrm{in}}^2)\big]. \label{eq22}
\eea
On the other hand, just inside the horizon, integrating $r_\ast$ yields
\be
(r_\ast)_{\textrm{in}}=\int \frac{dr}{f(r)}=\frac{1}{f'(r_+)}\ln \Big[\frac{(r_+-r)}{f'(r_+)-(r_+-r)f''(r_+)/2}\Big]. \label{eq23}
\ee
Substituting (\ref{eq23}) into (\ref{eq22}), we observe, just inside the horizon, the metric in the Kruskal coordinate system has the same form as (\ref{eq18}), only replacing the Kruskal coordinates outside the horizon ($T_{\textrm{out}}, X_{\textrm{out}}$) with that inside the horizon ($T_{\textrm{in}}, X_{\textrm{in}}$). Here, the Kruskal coordinates ($T_{\textrm{in}}, X_{\textrm{in}} $) which are valid inside the horizon, are written as
\bea
&&T_{\textrm{in}}=e^{\kappa_+(r_\ast)_{\textrm{in}}}\cosh(\kappa_+ \hat{t}_{\textrm{in}}),\nonumber\\
&&X_{\textrm{in}}=e^{\kappa_+(r_\ast)_{\textrm{in}}}\sinh(\kappa_+ \hat{t}_{\textrm{in}}).\label{eq24}
\eea
In short, in the Kruskal coordinates defined by (\ref{eq19}) and (\ref{eq24}), there is no spacetime singularity at the horizon, and the metric just outside and inside the horizon shares the same forms.  This two sets of the Kruskal coordinates are connected with
\begin{equation}
\hat{t}_{\textrm{in}}\rightarrow \hat{t}_{\textrm{out}}-i\frac{\pi}{2\kappa_+}, ~~(r_\ast)_{\textrm{in}}\rightarrow (r_\ast)_{\textrm{out}}+i\frac{\pi}{2\kappa_+},\label{eq25}
\end{equation}
under which, we have $T_{\textrm{in}}\rightarrow T_{\textrm{out}}$ and $X_{\textrm{in}}\rightarrow X_{\textrm{out}}$. Now, following the definition (\ref{eq10}), we find the relations connecting the null coordinates defined inside and outside the horizon
\begin{eqnarray}
u_{\textrm{in}}&=&\hat{t}_{\textrm{in}}-(r_\ast)_{\textrm{in}}\rightarrow \hat{t}_{\textrm{out}}-(r_\ast)_{\textrm{out}}-i\frac{\pi}{\kappa_+}\nonumber\\
v_{\textrm{in}}&=&\hat{t}_{\textrm{in}}+(r_\ast)_{\textrm{in}}\rightarrow  \hat{t}_{\textrm{out}}+(r_\ast)_{\textrm{out}}. \label{eq26}
\end{eqnarray}
Under this mapping, the right and left moving modes defined just inside and outside the horizon are connected by
\begin{eqnarray}
&&\Phi_{\textrm{in}}^{R}\rightarrow e^{-{\pi(\omega-\omega_0)}/{\kappa_+}}\Phi_{\textrm{out}}^{R}, \nonumber\\
&&\Phi_{\textrm{in}}^{L}\rightarrow \Phi_{\textrm{out}}^{L}.\label{eq27}
\end{eqnarray}
With this connection, the probability for the left moving mode to cross the horizon from inside is
\begin{equation}
\Gamma^{L}=|{\Phi_{\textrm{in}}^{L}}|^2\rightarrow |{\Phi_{\textrm{out}}^{L}}|^2=1.\label{eq28}
\end{equation}
This shows that the left moving (ingoing) mode is trapped inside the black hole, as expected. On the other hand, the right moving mode tunnels through the horizon and its probability, to go outside the horizon, as measured by an external observer is
\begin{equation}
\Gamma^{R}=|\Phi_{\textrm{in}}^{R}|^2\rightarrow |e^{-{\pi(\omega-\omega_0)}/{\kappa_+}}\Phi_{\textrm{out}}^{R}|^2=e^{-{2\pi(\omega-\omega_0)}/{\kappa_+}}.\label{eq29}
\end{equation}
Now, the total tunneling rate is given by $\Gamma=\Gamma^{L}\Gamma^{R}=e^{-{2\pi(\omega-\omega_0)}/{\kappa_+}}$, which could also be found in the usual approaches  \cite{tunneling1,t2,t3,t4,t5,t6}. Hence, under the Kruskal extension (\ref{eq19}) and (\ref{eq24}) introduced by us, a perfect tunneling picture has been provided during the charged particle's tunneling across the horizon. As a result, the left moving (ingoing) mode is expectedly trapped inside the black hole, and the right moving (outgoing) mode can quantum tunnel across horizon with the tunneling rate $\Gamma^{R}=e^{-{2\pi(\omega-\omega_0)}/{\kappa_+}}$. However, the Kruskal extension (\ref{eq1}) introduced by Zhao etc. fails to provide such a suitable tunneling picture for the charged particle's emission. In this picture, the ingoing mode is classically forbidden traveling towards the center of black hole, but can quantum tunnel across the horizon with the tunneling rate $\Gamma^{L}=e^{-\pi \omega_0/\kappa_+}$. And, the right moving mode is tunneling across the horizon with the probability $\Gamma^{R}=e^{-{\pi(2\omega-\omega_0)}/{\kappa_+}}$. This observation obviously violates the fact that the left moving (ingoing) mode is expected to trap inside the black hole, and its probability to go inside, as measured by an external observer, is to be unity. Also, the total tunneling rate $\Gamma=\Gamma^{L}\Gamma^{R}=e^{-{2\pi\omega}/{\kappa_+}}$ is problematic for describing the charged particle's tunneling. In the next section, with the aid of the Kruskal extension introduced by us, we develop the Banerjee-Majhi's treatment to the case of a charged black hole. More specifically, we will present the Hawking emission spectrum and entropy spectroscopy as tunneling from the horizon of the (charged) Reissner-Nordstr\"{o}m black hole.

\section{Quantum tunneling and emission spectrum} \label{sec2}
Under the Kruskal extension (\ref{eq19}) and (\ref{eq24}) introduced by us, the left and right moving mode defined inside and outside the horizon are related by (\ref{eq27}). In this section, with the aid of the relation (\ref{eq27}), we attempt to extend the Banerjee-Majhi's treatment to find the Hawking emission spectrum of the Reissner-Nordstr\"{o}m black hole. In doing so, one should find the density matrix for the observer, as stated in \cite{c1}. Before that, it is necessary to construct the physical state of the system, observed outside the horizon, for the $n$ number of non-interacting virtual pairs that are created inside the horizon as
\bea
|\Psi\rangle &=&N\sum_n|n_{\textrm{in}}^L\rangle\otimes |n_{\textrm{in}}^R\rangle \nonumber\\
&=&N\sum_n e^{-{\pi n(\omega-\omega_0)}/{\kappa_+}}|n_{\textrm{out}}^L\rangle\otimes |n_{\textrm{out}}^R\rangle,\label{eq31}
\eea
where $N$ is a normalization constant, which can be determined by the normalization condition $\langle\Psi|\Psi\rangle=1$. As a result, for bosons $N_{\textrm{boson}}=(1-e^{-{2\pi (\omega-\omega_0)}/{\kappa_+}})^{\frac{1}{2}}$, whereas for fermions $N_{\textrm{fermion}}=(1+e^{-{2\pi (\omega-\omega_0)}/{\kappa_+}})^{-\frac{1}{2}}$. In the following analysis, we only consider the boson case without loss of generality, since for fermions the analysis is identical. For bosons, the density operator can be constructed as
\begin{eqnarray}
\widehat{\rho}_{\textrm{boson}}&=&|\Psi\rangle_{\textrm{boson}}\langle\Psi|_{\textrm{boson}}\nonumber\\
&=& (1-e^{-{2\pi (\omega-\omega_0)}/{\kappa_+}})\sum_{n,m} e^{-{\pi n(\omega-\omega_0)}/{\kappa_+}}\nonumber\\
&\times& e^{-{\pi m(\omega-\omega_0)}/{\kappa_+}}|n_{\textrm{in}}^L\rangle\otimes |n_{\textrm{in}}^R\rangle \langle m_{\textrm{in}}^R|\otimes \langle m_{\textrm{in}}^L|. \label{eq32}
\end{eqnarray}
As mentioned in Sec. \ref{sec1}, the left moving modes are all trapped inside the horizon. Now, tracing out all such left modes, we find the reduced density operator for the right moving modes, is given by
\bea
\widehat{\rho}_{\textrm{boson}}^R&=&(1-e^{-{2\pi (\omega-\omega_0)}/{\kappa_+}}) \nonumber\\
&\times& \sum_n e^{-{2\pi n(\omega-\omega_0)}/{\kappa_+}}|n_{\textrm{in}}^R\rangle \langle n_{\textrm{in}}^R|.\label{eq33}
\eea
Therefore, the average number of particles, which is detected by an observer living outside the horizon, is read off
\begin{equation}
\langle n\rangle_{\textrm{boson}}=\textrm{trace}(\widehat{n}\widehat{\rho}_{\textrm{boson}}^R)=\big(e^{2\pi (\omega-\omega_0)/\kappa_+}-1\big)^{-1}.\label{eq34}
\end{equation}
This distribution for boson corresponds to a blackbody emission spectrum with a temperature $T_+=\frac{\kappa_+}{2\pi}$, which is consistent with the Hawking's observation. Also, we can turn to other methods to reproduce the same distribution for boson, such as the generalized tortoise coordinate transformation (GTCT) etc.
Hence, under the Kruskal extension introduced by us, the Hawking emission spectrum of the Reissner-Nordstr\"{o}m black hole can be well reproduced in the Banerjee-Majhi's treatment.

\section{Quantum tunneling and entropy quantum} \label{sec3}
In this section, under the relation (\ref{eq27}), we aim to produce the entropy spectrum as tunneling from the horizon of the Reissner-Nordstr\"{o}m black hole in the Banerjee-Majhi's treatment. As mentioned in Sec. \ref{sec1}, particle creates just inside the horizon, and the left moving mode is trapped inside the horizon while the right moving mode can tunnel across the horizon to be observed by an outside observer. Therefore, the average value of the energy $E=\omega-\omega_0$ can be found as
\begin{eqnarray}
\langle E\rangle &=&\frac{\int_{0}^{\infty} (\Phi_{\textrm{in}}^R)^\ast E \Phi_{\textrm{in}}^R d E }{\int_{0}^{\infty} (\Phi_{\textrm{in}}^R)^\ast  \Phi_{\textrm{in}}^R d E}\nonumber\\
&=&\frac{\int_{0}^{\infty} e^{-\frac{\pi E}{\kappa_+}}(\Phi_{\textrm{out}}^R)^\ast E e^{-\frac{\pi E}{\kappa_+}}\Phi_{\textrm{out}}^R d E }{\int_{0}^{\infty} e^{-\frac{\pi E}{\kappa_+}}(\Phi_{\textrm{out}}^R)^\ast  e^{-\frac{\pi E}{\kappa_+}}\Phi_{\textrm{out}}^R d E}\nonumber\\
&=& \frac{\int_{0}^{\infty} E e^{-\frac{E}{T_+}}d E}{\int_{0}^{\infty}  e^{-\frac{E}{T_+}}d E}=T_+. \label{eq41}
\end{eqnarray}
Here, to compute this expression it is important to recall that the observer is located outside the
horizon. Hence it is essential to recast the ``in" expressions into their corresponding ``out" expressions using the relation (\ref{eq27}) and then perform the integrations. Proceeding in a similar manner, one can reproduce the average squared energy of the particle detected by the observer living outside the horizon, which is given by
\begin{eqnarray}
\langle E^2\rangle &=&\frac{\int_{0}^{\infty} (\Phi_{\textrm{in}}^R)^\ast E^2 \Phi_{\textrm{in}}^R d E }{\int_{0}^{\infty} (\Phi_{\textrm{in}}^R)^\ast  \Phi_{\textrm{in}}^R d E}\nonumber\\
&=&\frac{\int_{0}^{\infty} e^{-\frac{\pi E}{\kappa_+}}(\Phi_{\textrm{out}}^R)^\ast E^2 e^{-\frac{\pi E}{\kappa_+}}\Phi_{\textrm{out}}^R d E}{\int_{0}^{\infty} e^{-\frac{\pi E}{\kappa_+}}(\Phi_{\textrm{out}}^R)^\ast  e^{-\frac{\pi E}{\kappa_+}}\Phi_{\textrm{out}}^R d E}\nonumber\\
&=& \frac{\int_{0}^{\infty} E^2 e^{-\frac{E}{T_+}}d E}{\int_{0}^{\infty}  e^{-\frac{E}{T_+}}d E}=2T_+^2. \label{eq42}
\end{eqnarray}
With the aid of (\ref{eq41}) and (\ref{eq42}), one can find the uncertainty of the energy $E$ detected by the observer living outside the horizon, is given by
\begin{equation}
\Delta E=\sqrt{\langle E^2\rangle-\langle E\rangle^2}=T_+.\label{eq43}
\end{equation}
In view of (\ref{eq43}), the uncertainty in the characteristic frequency of the tunneling mode is $\Delta f={\Delta E}={T_+}$. On the other hand, the energy uncertainty can be treated as the lack of information in energy of the black hole during the particle emission. Also, in the information theory,
the entropy acts as the lack of information. To connect these two quantities, we can use the first law of thermodynamic as $\Delta E=T_+\Delta S_{\textrm{ch}}$. According to the Bohr-Sommerfeld quantization rule, substituting this uncertainty into the first law of thermodynamic yields the entropy spectrum as
\begin{equation}
 S_{\textrm{h}}=n.\label{eq44}
\end{equation}
Obviously, $\Delta S_{\textrm{h}}=(n+1)-n=1$, which shows that the entropy of the Reissner-Nordstr\"{o}m black hole is quantized in units of the identity. It is of interest that this entropy spectrum is exactly identical with the result of Hod by considering the Heisenberg uncertainty principle and Scwinger-type charge emission process \cite{hod}. Hence, under the suitable Kruskal extension (\ref{eq19}) and (\ref{eq24}) introduced by us, the entropy spectrum of the (charged) Reissner-Nordstr\"{o}m  black hole can well be reproduced in the Banerjee-Majhi's treatment. On the other hand, this entropy spectrum is consistent with that in \cite{RBE,BRM,JHC,JHC1} for black hole with no charge, which shows the entropy quantum is independent of the black hole parameters. This is a desiring result for the forthcoming quantum gravity theory. It should be noted that the entropy quantum in our analysis (combined with statistical physics, quantum mechanics and black hole physics) is on a semiclassical level, but it is a much better approximation since a self-consistent theory of quantum gravitation which give a definite answer to the entropy quantum is lacking.

\section{Conclusion and Discussion}\label{sec4}
In this paper, we attempt to develop the Banerjee-Majhi's treatment to the case of the (charged) Reissner-Nordstr\"{o}m black hole. First, we revisit the Kruskal extension introduced by Zhao etc. in \cite{zhaor}, and analyze its induced \emph{unphysical} phenomenon.
With this in mind, a suitable Kruskal extension for the charged black hole is then introduced by us to provide a perfect tunneling picture during the charged particle's emission. Finally, under the new Kruskal extension, we revisit the Hawking emission spectrum and entropy spectroscopy as tunneling from the Reissner-Nordstr\"{o}m black hole. The result shows that the tunneling method is so universally robust that the Hawking blackbody emission spectrum from a charged black hole can be well reproduced in the tunneling mechanism, and its induced entropy quantum is a much better approximation for the forthcoming quantum gravity theory.

In our analysis, the entropy quantum is obtained from the combination of statistical physics, quantum mechanics and black hole physics. Hence, our result is on a semiclassical level, but it is a much better approximation since a self-consistent theory of quantum gravitation which give a definite answer to the entropy quantum is lacking. On the other hand, our analysis in this paper is applicable for a variety of a charged or rotating black hole since after a dimensional reduction technique near the horizon the effective two-dimensional theory for a charged or rotating black hole is described in the same forms (For detailed information, please refer to the Hawking radiation via gauge and gravitational anomalies). In this sense, our analysis is universal.

\section*{Acknowledgments}
This work is supported by the National Natural Science Foundation of China with Grant No.
11005086, and by the Sichuan Youth Science and Technology
Foundation with Grant No. 2011JQ0019, and by a starting fund of China West Normal University with Grant No. 10B016.

%

\end{document}